\documentclass[a4paper,11pt]{article}
\pdfoutput=1 
\usepackage{jheppub} 
\usepackage{epsfig}
\usepackage{caption}
\usepackage{subcaption}
\usepackage{color,soul}
\usepackage{float}
\usepackage{amsfonts}
\usepackage{url}
\usepackage{slashed}
\usepackage{accents}
\usepackage{lipsum}
\usepackage{mathtools}
\usepackage{physics}

\usepackage[normalem]{ulem}

\hypersetup{
  bookmarks=true,         % show bookmarks bar?
  unicode=false,          % non-Latin characters in Acrobat?s bookmarks
  pdftoolbar=true,        % show Acrobat?s toolbar?
 pdfmenubar=true,        % show Acrobat?s menu?
 pdffitwindow=true,     % window fit to page when opened
 pdfstartview={FitH},    % fits the width of the page to the window
 pdfsubject={Dark Matter},   % subject of the document
 pdfnewwindow=true,      % links in new window
 pdfcreator={RevTeX},
 colorlinks=true,       % false: boxed links; true: colored links
 linkcolor=red,          % color of internal links
 citecolor=blue,        % color of links to bibliography
 filecolor=black,      % color of file links
 urlcolor=blue,           % color of external links
  }

\title{Momentum distribution of dark matter produced in inflaton decay: effect of inflaton mediated scatterings}
\author[]{Avirup Ghosh}
\author[]{and Satyanarayan Mukhopadhyay}
\affiliation[]{School of Physical Sciences, Indian Association for the Cultivation of Science, 2A and 2B Raja S.C. Mullick Road, Kolkata 700 032}

\emailAdd{spsag2510@iacs.res.in}
\emailAdd{tpsnm@iacs.res.in}

\abstract{Post-inflationary reheating is a widely discussed mechanism for non-thermal production of dark matter (DM). In this scenario the momentum distribution of the produced DM particles is usually taken to be the one obtained at reheating, red-shifted at later times due to the expansion of the Universe. However, since in such a scenario both the DM and the standard model (SM) fields couple to the inflaton, the DM particles necessarily undergo self-scatterings, as well as elastic and inelastic scattering reactions with the SM bath, all of which proceed through $s-$channel or $t-$channel inflaton exchange. We compute the momentum distribution of the DM particles including the effect of these scatterings, and find that the distributions can be significantly altered, even though  DM remains non-thermal throughout the cosmological evolution. We observe that if the inflaton dominantly couples to the SM Higgs boson through a renormalizable interaction, then reheating temperatures and inflaton masses at the TeV scale lead to a large effect from the scattering processes, with the DM-inflaton coupling constrained by the DM density. The scattering effects are found to be sensitive to the duration of the reheating process -- larger the duration, more momentum modes are filled at reheating, leading to an enhanced scattering probability. We also obtain the free-streaming length of such DM using the resulting non-thermal momentum distribution, which can be used to estimate the implications of the Lyman-$\alpha$ constraints on the DM mass. It is observed that in the scenarios considered, including the scattering effects can reduce the DM average velocity at matter-radiation equality, and its free-streaming length, by upto a factor of $40$, thereby making the constraints on light DM produced in inflaton decay significantly weaker.}

\begin{document} 
\maketitle
\flushbottom  
\section{Introduction}
\label{sec:sec1}
Inflation is a leading candidate for a theory of the pre-radiation dominated Universe, in which the horizon and flatness problems in big-bang cosmology can be addressed, and which also makes predictions about properties of fluctuations in the cosmic microwave background and large-scale structure~\cite{Weinberg:2008zzc,Starobinsky:1979ty, Starobinsky:1980te, Kazanas:1980tx, Guth:1980zm,Linde:1981mu,Linde:1982uu,Albrecht:1982wi}.

At the end of the slow-roll phase of inflation, the inflaton field $\phi$ undergoes a damped oscillation around the minimum of the potential $\phi_0$. When $\phi$ is close to $\phi_0$, the potential can be approximated by a quadratic form:
\begin{equation}
V(\phi) = \frac{1}{2}m^2_\phi\,(\phi - \phi_0)^2 .
\label{eqn:infpot}
\end{equation}  
This is equivalent to the field theory of spin-0 particles of mass $m_\phi$, and negligible velocity. Furthermore, since inflation needs to end eventually, giving rise to  a radiation dominated Universe, the inflaton field must also couple and decay to other fields including standard model (SM) matter and radiation. This epoch of inflaton oscillation and decay is known as the period of reheating~\cite{Weinberg:2008zzc,Linde:1981mu,Albrecht:1982mp,Dolgov:1982th,Abbott:1982hn,  Dolgov:1989us, Traschen:1990sw,Kofman:1994rk,Kofman:1997yn,Allahverdi:2010xz}. If the dark matter (DM) field couples to the inflaton as well, DM may be produced non-thermally during the reheating epoch. Such a possibility has been extensively considered in the literature, see for example, Refs~\cite{Moroi:1994rs, Kawasaki:1995cy, Moroi:1999zb, Chung:1998zb, Chung:1998ua, Jeong:2011sg, Ellis:2015jpg, Harigaya:2014waa, Drees:2021lbm, Garcia:2018wtq, Harigaya:2019tzu, Garcia:2020eof,  Garcia:2020wiy, Moroi:2020has, Bhattacharya:2020zap, Haque:2022kez, Haque:2021mab}. Light, cold DM may also be produced during reheating, and with suitable choices of the reheating temperature $T_R$ and the inflaton mass $m_\phi$, the DM particles from inflaton decay can become cold enough due to the red-shift of the momenta~\cite{Moroi:2020has}.

Since the DM field in this scenario couples to the inflaton, inflaton mediated DM self-scatterings necessarily take place. In addition, since the SM fields also couple to the inflaton, DM-SM scattering processes become relevant as well. These scattering reactions, if they have appreciable rates, can modify the momentum distribution of the DM particles. Furthermore, once the inflaton mass and its couplings to the SM and DM fields are fixed to reproduce a given reheating temperature and a desired DM density, the DM-DM and DM-SM scattering rates are completely determined. The impact of the scattering processes on the DM momentum distribution should therefore be studied in detail. This is the primary objective of the present study. For light DM, the momentum distribution is directly relevant in determining the constraints from structure formation. We have, to this end, made a simple estimate of the free-streaming length of the DM particles, which can be used to bound the DM mass using constraints from the Lyman-$\alpha$ forest data~\cite{Kolb:1990vq, Boyarsky:2008xj, Irsic:2017ixq}. This work, therefore, should be relevant for more accurate determinations of the structure formation constraints on light non-thermal DM produced in inflaton decay.

The rest of the paper is organized as follows. In Sec.~\ref{sec:sec2} we discuss the production of DM and SM radiation during reheating, and obtain the DM number density and momentum distribution at the end of the reheating epoch, including the effects of quantum statistics and back-reaction. We then describe the framework adopted to obtain our primary new results on the effect of inflaton-mediated scatterings on the DM momentum distribution in Sec.~\ref{sec:sec3}. The computations of the DM average velocity and its co-moving free-streaming length are discussed in Sec.~\ref{sec:sec4}. We present our results on the modifications to the DM momentum distribution due to scatterings, and the impact of the reheating temperature, inflaton mass, the DM-inflaton coupling and the duration of the reheating process on it in Sec.~\ref{sec:sec5}. We also discuss the impact of the scatterings on the resulting DM average velocity at matter-radiation equality and the DM free-streaming length, along with the implications of the Lyman-$\alpha$ constraints in this section. Sec.~\ref{sec:sec6} summarizes our results.

\section{Dark matter production during reheating}
\label{sec:sec2}
We consider a minimal scenario in which the scalar inflaton field $\phi$ couples through renormalizable interactions to both the SM and the DM sectors. We take the DM field $\psi$ to be an SM singlet fermion in the following discussion. The results can be straightforwardly extended to the case of a singlet scalar DM as well, with the essential modification being in the quantum statistics involved. The relevant interaction Lagrangian is given by 
\begin{equation}
\mathcal{L}_{\rm int} \supseteq - \mu_{H}  H^\dagger H \phi - \lambda \overline{\psi}\psi \phi .
\label{eqn:lag}
\end{equation}
Here, $H$ is the SM Higgs doublet, the only SM field to which the inflaton has a renormalizable coupling~\footnote{The scenario considered here is a minimal one with the inflaton coupling to only the Higgs field through a renormalizable interaction. In general, the inflaton can couple to all the SM fields, and the broad qualitative conclusions in the subsequent sections of this study should continue to hold in such scenarios as well. Similar considerations apply to the $R^2$ (Starobinsky) inflation scenario, in which the reheating in the SM sector proceeds through the particle production by the curvature scalar, which is universally coupled to all types of elementary particles.}. Although here we have considered the reheating of the SM sector through the trilinear scalar interaction, the quartic interaction term $\phi^2 H^\dagger H$ can also be relevant~\cite{Lebedev:2021tas}. 

\subsection{DM number density and distribution in the instantaneous decay approximation}

To understand the parametric dependence of the DM number density and its momentum distribution on the reheating process variables, we can start with the approximation that the inflaton field decays instantaneously, giving rise to a radiation dominated Universe. In that case, the DM number density at the reheating time $t_R$, $n_\psi(t_R)$, can be computed as follows. Since the DM is produced from the decay of the inflaton quanta, $\phi \rightarrow \psi \psi$, we have
\begin{equation}
n_\psi(t_R) \simeq 2 \times {\rm BR}(\phi \rightarrow \psi) n_\phi(t_R),
\label{eqn:DMno1}
\end{equation}
where, we have added up the contributions to the DM particle and anti-particle number densities in $n_\psi(t_R)$, and ${\rm BR}(\phi \rightarrow \psi)$ denotes the branching ratio of $\phi$ decays to a DM pair. We also have the inflaton number density $n_\phi(t_R) = \dfrac{\rho_\phi(t_R)}{m_\phi}$, where $\rho_\phi(t_R)$ is the inflaton energy density at $t_R$. Let us also denote by $\Gamma_\phi$, $\Gamma_R$ and $\Gamma_\psi$ the total decay width of the inflaton, and its partial decay widths to the SM radiation and DM, respectively. In terms of the couplings in Eq.~\ref{eqn:lag}, we have $\Gamma_R = \mu_H^2/(32 \pi m_\phi)$ and $\Gamma_\psi=\lambda^2 m_\phi/(8 \pi)$.
Assuming that the inflaton dominantly decays to the SM sector (which is necessary in this framework to have a radiation-dominated Universe at the big-bang nucleosynthesis (BBN) epoch), we can approximately write, $\Gamma_\phi \simeq \Gamma_R$, and $\rho_\phi(t_R) \simeq \rho_R(t_R)$, where $\rho_R(t_R)$ is the radiation energy density at $t_R$. Furthermore, the reheating time $t_R$ can be defined by the following relation
\begin{equation}
\Gamma_\phi \simeq H(t_R) \simeq \frac{\sqrt{\rho_R(t_R)}}{\sqrt{3}M_{\rm Pl}} ,
\label{eqn:phiwidth}
\end{equation}
where, $H(t_R)$ is the Hubble expansion rate at $t_R$ and we have defined the reduced Planck mass by $M_{\rm Pl}^2 = 1/8 \pi G$. With this, Eq.~\ref{eqn:DMno1} becomes 
\begin{equation}
n_\psi(t_R) \simeq \frac{2\Gamma_\psi}{H(t_R)}\,\frac{\rho_R(t_R)}{m_\phi}.
\label{eqn:DMno1A}
\end{equation}
Further assuming that the SM radiation bath thermalizes instantaneously with a temperature $T_R$ at time $t_R$, we have
\begin{equation}
n_\psi(t_R) \simeq \frac{2\sqrt{3}\Gamma_\psi\,M_{\rm Pl}}{m_\phi}\left(\frac{\pi^2\,g_*(T_R)}{30} \right)^{1/2}\,T^2_R,
\label{eqn:DMno2}
\end{equation}
where, $g_*(T_R)$ is the number of relativistic degrees of freedom at the temperature $T_R$.
Thus the DM number density at $t=t_R$ is proportional to the combination $T_R^2/m_\phi$, in addition to $\Gamma_\psi$. If no DM particle is produced or annihilated after $t=t_R$, $n_\psi(t_R)$ can be used to obtain the DM density $\Omega_\psi = \rho_\psi (t_0)/\rho_c$ at the present epoch $t=t_0$, where $\rho_c$ is the critical density:
\begin{equation}
\Omega_\psi =\frac{9\sqrt{10}}{2 \pi} \frac{m_\psi\,s_0\,\Gamma_\psi\,M_{\rm Pl}}{\rho_c\,m_\phi\,T_R \sqrt{g_*(T_R)}}.
\label{eqn:DMrelic1}
\end{equation}
Here, $s_0$ is the entropy density at the present epoch. Thus, keeping the partial decay width $\Gamma_\psi$ fixed, $\Omega_\psi$ is decreased for higher values of $T_R m_\phi$. 

Having obtained the DM number density $n_\psi(t_R)$, we can also determine its momentum distribution $f_\psi(\vec{k},t_R)$ using the same instantaneous decay approximation, where the distribution function is normalized as 
\begin{equation}
n_\psi(t) = g_\psi \int \frac{d^3k}{(2\pi)^3}\,f_\psi(\vec{k},t),
\label{eqn:numberint}
\end{equation}
$g_\psi$ being the internal degrees of freedom of $\psi$. At $t=t_R$, all the DM particles are produced with momentum $k = p_\psi \simeq m_\phi/2$, for $m_\phi >> m_\psi$. Hence, 
\begin{equation}
f_\psi(k,t_R) = A\,\delta(k-p_\psi),
\label{eqn:fdist1}
\end{equation}
where, $A$ is a proportionality constant and by isotropy of the Universe $f_\psi$ is a function of $|\vec{k}| \equiv k$ only. The constant $A$ can be determined using Eq.~\ref{eqn:numberint} with the $n_\psi (t_R)$ obtained in Eq.~\ref{eqn:DMno1A} leading to
\begin{equation}
f_\psi(k,t_R) = \frac{4\pi^2}{g_\psi\,p^2_\psi}\,\frac{{\rm BR}(\phi \rightarrow \psi)}{m_\phi}\,\rho_R(t_R)\delta(k-p_\psi).
\label{eqn:fdist2}
\end{equation}
In the absence of any collisions, for $t>t_R$, the momenta simply redshift as $k(t) \propto 1/a(t)$, where $a(t)$ is the scale factor in the Friedmann-Robertson-Walker space-time metric. 

\subsection{Beyond the instantaneous decay approximation}
To obtain a more accurate expression for the number density of the DM $n_\psi (t)$, we need to consider the evolution of the energy densities of the inflaton field ($\rho_\phi$) and the SM radiation bath ($\rho_R$), and $n_\psi$, which are governed by the following coupled set of approximate equations,
\begin{align}
\frac{d\rho_\phi}{dt}+3\,H\,\rho_\phi & = -\Gamma_\phi\,\rho_\phi, \label{eqn:evoleqns1} \\
\frac{d\rho_R}{dt}+4\,H\,\rho_R & = \Gamma_R\,\rho_\phi, \label{eqn:evoleqns2} \\
\frac{dn_\psi}{dt}+3\,H\,n_\psi & = 2\,\Gamma_\psi\,\frac{\rho_\phi}{m_\phi}.
\label{eqn:evoleqns3} 
\end{align}
along with the Friedmann equation for the Hubble expansion rate $H$
\begin{equation}
3 M_{\rm Pl}^2 H^2  = \rho_R + \rho_\phi + \rho_\psi.
\label{eqn:hubble}
\end{equation}
In writing these equations, we have made the approximation that the back-reactions and the effect of quantum statistics are not significant. However, they will be important for the light DM to be considered here, and will be included in the subsequent discussion of the momentum distribution and number density of DM. We have also considered the DM to be a massive matter particle, whose number in a co-moving volume ($n_\psi a^3$) is conserved in the absence of production and annihilation processes. The factor of $2$ in the RHS of Eq.~\ref{eqn:evoleqns3} accounts for the fact that $2$ DM particles are produced in the decay of each inflaton.

Eq.~\ref{eqn:evoleqns1} for $\rho_\phi$ can be integrated to obtain 
\begin{equation}
\rho_\phi(t) = \rho_\phi(t_I)\,\left(\frac{a(t_I)}{a(t)}\right)^3\,e^{-\Gamma_\phi(t-t_I)},
\label{eqn:infdense}
\end{equation}
where, the time $t_I$ is the beginning of the period of inflaton oscillations and decay, when the energy density in the inflaton field is $\rho_\phi(t_I)$. We can also integrate Eq.~\ref{eqn:evoleqns3} using the solution in Eq.~\ref{eqn:infdense} as an input, and obtain
\begin{equation}
n_\psi(t) \simeq 2 \left(\frac{\rho_\phi(t_I)}{m_\phi}\right)\,\left(\frac{a(t_I)}{a(t)}\right)^3\,{\rm BR}(\phi \rightarrow \psi)\,\left[1-e^{-\Gamma_\phi(t-t_I)}\right].
\label{eqn:DMno3}
\end{equation}
We can further rewrite Eq.~\ref{eqn:DMno3} with the help of Eq.~\ref{eqn:infdense} to express $n_\psi(t_R)$ as follows:
\begin{equation}
n_\psi(t_R) \simeq 2\,{\rm BR}(\phi \rightarrow \psi)\,n_\phi(t_R)\,\left[e^{\Gamma_\phi(t_R-t_I)}-1\right].
\label{eqn:numberint1}
\end{equation}
Comparing Eq.~\ref{eqn:numberint1} with Eq.~\ref{eqn:DMno1}, we see that compared to the instantaneous decay approximation, there is an enhancement in $n_\psi(t_R)$, due to a larger inflaton number density $n_\phi (t)$, for $t_I \leq t \leq t_R$. With $\Gamma_\phi t_R \sim 1$, and $\Gamma_\phi t_I \rightarrow 0$, this amounts to an enhancement by a factor of around $(e-1) \simeq 1.72$.

Finally, Eq.~\ref{eqn:evoleqns2} gives the radiation energy density as
\begin{equation}
\rho_R(t) \simeq \rho_\phi(t_I) \Gamma_R \frac{a^3(t_I)}{a^4(t)}\,\int_{t_I}^{t}\,dt^\prime\,a(t^\prime)\,e^{-\Gamma_\phi(t^\prime-t_I)}.
\label{eqn:raddense}
\end{equation}
In writing Eqs.~\ref{eqn:DMno3} and ~\ref{eqn:raddense}, we have used the boundary conditions that $n_\psi(t_I)=0=\rho_R(t_I)$. By solving the coupled system of equations~\ref{eqn:evoleqns1}-\ref{eqn:hubble} numerically, we can determine $\rho_R(t)$. Simple approximate solutions may also be obtained by assuming that the energy density of the Universe is dominated by a stable matter field $\phi$ during the time $t_I \leq t < t_R$, where $t_R$ is the time when radiation domination starts. Using the fact that in a stable matter dominated Universe $H=2/3t$, we obtain at $t=t_I$
\begin{equation}
\rho_\phi(t_I) \simeq \frac{4M^2_{\rm Pl}}{3\,t^2_I},
\label{eqn:infdenseini}
\end{equation}
and hence,
\begin{equation}
\rho_\phi(t) \simeq \frac{4M^2_{\rm Pl}}{3\,t^2}e^{-\Gamma_\phi(t-t_I)}\,\,\,{\rm for}\,\,\,t_I \leq t \leq t_R,
\label{eqn:infdense2}
\end{equation}
where we have used $a(t) \propto t^{2/3}$ in a stable matter dominated era. Using Eq.~\ref{eqn:infdense2} in Eq.~\ref{eqn:evoleqns2} we can solve for $\rho_R(t)$ in terms of an infinite series as~\cite{Bhatia:2020itt}
\begin{eqnarray}
\rho_R(t) &=& \frac{4M^2_{\rm Pl}\Gamma^2_\phi}{3}e^{\Gamma_\phi\,t_I}\, \bigg[\overset{\infty}{\underset{n=0}{\sum}}\frac{(-\Gamma_\phi\,t)^{n-1}}{n!(n+5/3)} - \left(\frac{t_I}{t}\right)^{8/3} \overset{\infty}{\underset{n=0}{\sum}}\frac{(-\Gamma_\phi\,t_I)^{n-1}}{n!(n+5/3)} \bigg].
\label{eqn:raddense2}
\end{eqnarray}
Here again, we have imposed the boundary condition that $\rho_R(t_I)=0$.

\subsection{Including the effects of back-reaction and quantum statistics}
The computation so far has not included the effect of back-reaction and quantum statistics. As the DM number density builds up, both the effects may become relevant. In order to include these effects let us consider the Boltzmann kinetic equation for $f_\psi(\vec{k},t)$, with $f_\psi(\vec{k},t) d^3k$ being the number density of DM particles in the momentum range $\vec{k}$ to $\vec{k}+d\vec{k}$:
\begin{equation}
\frac{\partial\,f_\psi(\vec{k},t)}{\partial\,t} - H\,\vec{k}.\vec{\nabla}_{\vec{k}}\,f_\psi(\vec{k},t) = f_\psi^{\rm coll}({\vec{k}},t),
\label{eqn:BEinf1}
\end{equation}
where, the collision term for the process $\phi(\vec{p}) \leftrightarrow \psi(\vec{k})+\psi(\vec{k^\prime})$ is given by
\begin{eqnarray}
f_\psi^{\rm coll}({\vec{k}},t) &=& -\frac{1}{2E_{\vec{k}}}\int \frac{d^3\vec{k}^\prime}{(2\pi)^3\,2E^\prime_{\vec{k^\prime}}}\frac{d^3\vec{p}}{(2\pi)^3\,2E_{\vec{p}}}(2\pi)^4\delta^4(p-k^\prime-k) \times \nonumber\\
&& 2\bigg[|M|^2_{\psi \psi \rightarrow \phi} f_\psi({\vec{k}},t)  f_\psi({\vec{k}^\prime},t)(1+f_\phi(\vec{p},t))\nonumber\\
&& 
-|M|^2_{\phi \rightarrow \psi \psi} f_\phi(\vec{p},t) (1- f_\psi({\vec{k}},t)) (1- f_\psi({\vec{k^\prime}},t))\bigg]. 
\label{eqn:BEinf1collA}
\end{eqnarray}
In writing this equation we have assumed that the distribution functions for the DM particle and anti-particle are the same. 

Since the inflaton field during reheating may be taken to be dominated by its zero mode, the decaying inflaton particles are nearly at rest. Thus, we can reduce Eq.~\ref{eqn:BEinf1collA} to the following form
\begin{equation}
f_\psi^{\rm coll}({\vec{k}},t) = 2 \Gamma_{\psi} n_\phi(t) \left(\frac{2\pi^2}{p^2_{\psi}}\right)\delta(|\vec{k}|-p_{\psi}) \left[(1- f_\psi({\vec{k}},t))(1 - f_\psi({-\vec{k}},t)) - f_\psi({\vec{k}},t) f_\psi({-\vec{k}},t) \right].
\label{eqn:BEinf1collB} 
\end{equation}
In deriving Eq.~\ref{eqn:BEinf1collB}, we have assumed that the phase-space density of the inflaton quanta is large, and hence $1+f_\phi(\vec{p},t) \simeq f_\phi(\vec{p},t)$. Here, as before, $p_\psi$ is the momentum of the DM particle at production, and is given by
\begin{equation}
p_{\psi} = \frac{m_\phi}{2} \sqrt{1-\frac{4m_\psi^2}{m_\phi^2}}.
\label{eqn:mtmtR}
\end{equation}
For $f_\psi({\vec{k}},t) << 1$, the effects of back-reaction and Fermi statistics are not important, and integrating Eq.~\ref{eqn:BEinf1} in that limit reproduces Eq.~\ref{eqn:DMno3}, as expected.

The DM momentum $\vec{k}$ at time $t$ is related to its production time $t_p$, where $t_I \leq t_p \leq t_R$, since at $t_p$ its momentum is always $p_\psi$, which is then redshifted to $\vec{k}$ at time $t$. Hence, the $\vec{k}$ dependence of $f_\psi({\vec{k}},t)$ may be replaced by its $t_p$ dependence as follows:
\begin{equation}
\frac{df_\psi(t_p,t)}{dt} = 2 \Gamma_{\psi} n_\phi(t) \left(\frac{2\pi^2}{p^2_{\psi}}\right)\frac{\delta(t-t_p)}{p_\psi\,H(t_p)}\,\left(1-2f_\psi(t_p,t)\right),
\label{eqn:BEinf2}
\end{equation}
Here, we have used the relation
\begin{equation}
t_p = t\left(\frac{|\vec{k}(t)|}{p_\psi}\right)^{3/2},
\label{eqn:prodtime}
\end{equation}
which is obtained for $a(t)\propto t^{2/3}$ valid in a stable matter dominated Universe. In writing the collision term in Eq.~\ref{eqn:BEinf2} from Eq.~\ref{eqn:BEinf1collB}, we have also imposed the condition of isotropy of the Universe, which implies $f_\psi(|\vec{k}|,t) = f_\psi(\vec{k},t) = f_\psi(-\vec{k},t)$.

Integrating Eq.~\ref{eqn:BEinf2} and imposing the boundary condition that $f_\psi({\vec{k}},t_I)=0$, where $t_I$ is the time reheating begins, we obtain
\begin{equation}
f_\psi(t_p,t) = \frac{1}{2}\left(1-e^{-2\bar{f}(t_p)}\right)\,\Theta(p_\psi - k(t))
\label{eqn:distFB}
\end{equation}
where, 
\begin{equation}
\bar{f}(t_p) = \frac{4\pi^2\,\Gamma_\psi}{p^3_\psi}\frac{\rho_\phi(t_p)}{m_\phi}\frac{1}{H(t_p)},
\label{eqn:distbasic}
\end{equation}
Here, the $\Theta-$function ensures the condition that $k(t)<p_\psi$.

It is useful to observe the difference between the distribution functions obtained (a) without back-reaction and Fermi statistics, (b) including back-reaction but with classical statistics, and (c) with both back-reaction and Fermi statistics. While scenario-(c) is already discussed above, the distribution in scenario-(a) is found to be
\begin{equation}
f_{\psi\,(a)}(t_p,t) = \bar{f}(t_p)\,\Theta(p_\psi - k(t)),
\label{eqn:distnoB}
\end{equation}
and in scenario-(b) the distribution is obtained as follows
\begin{equation}
f_{\psi\,(b)}(t_p,t) = \tanh\left(\bar{f}(t_p)\right)\,\Theta(p_\psi - k(t)).
\label{eqn:distB}
\end{equation}
The function $\bar{f}(t_p)$ in Eq.~\ref{eqn:distbasic} can be written in terms of the momentum $k(t_R)$ at $t=t_R$ as
\begin{equation}
\bar{f}(k(t_R)) \simeq \frac{4\pi^2\,{\rm BR}(\phi \rightarrow \psi)}{p^3_\psi}\,\frac{\rho_\phi(t_R)}{m_\phi}\left[\frac{3}{2} \frac{\exp(1-(k(t_R)/p_\psi)^{3/2})}{(k(t_R)/p_\psi)^{3/2}}\right],
\label{eqn:fbareq}
\end{equation}
where, we have set $\Gamma_\phi t_R \sim 1$. The distribution function in Eq.~\ref{eqn:distFB}, along with Eq.~\ref{eqn:fbareq} are used as the initial condition at $t=t_R$ for the evolution of $f_\psi(\vec{k},t)$ due to the effect of collisions.

\begin{figure} [htb!]
\begin{center} 
\includegraphics[scale=0.245]{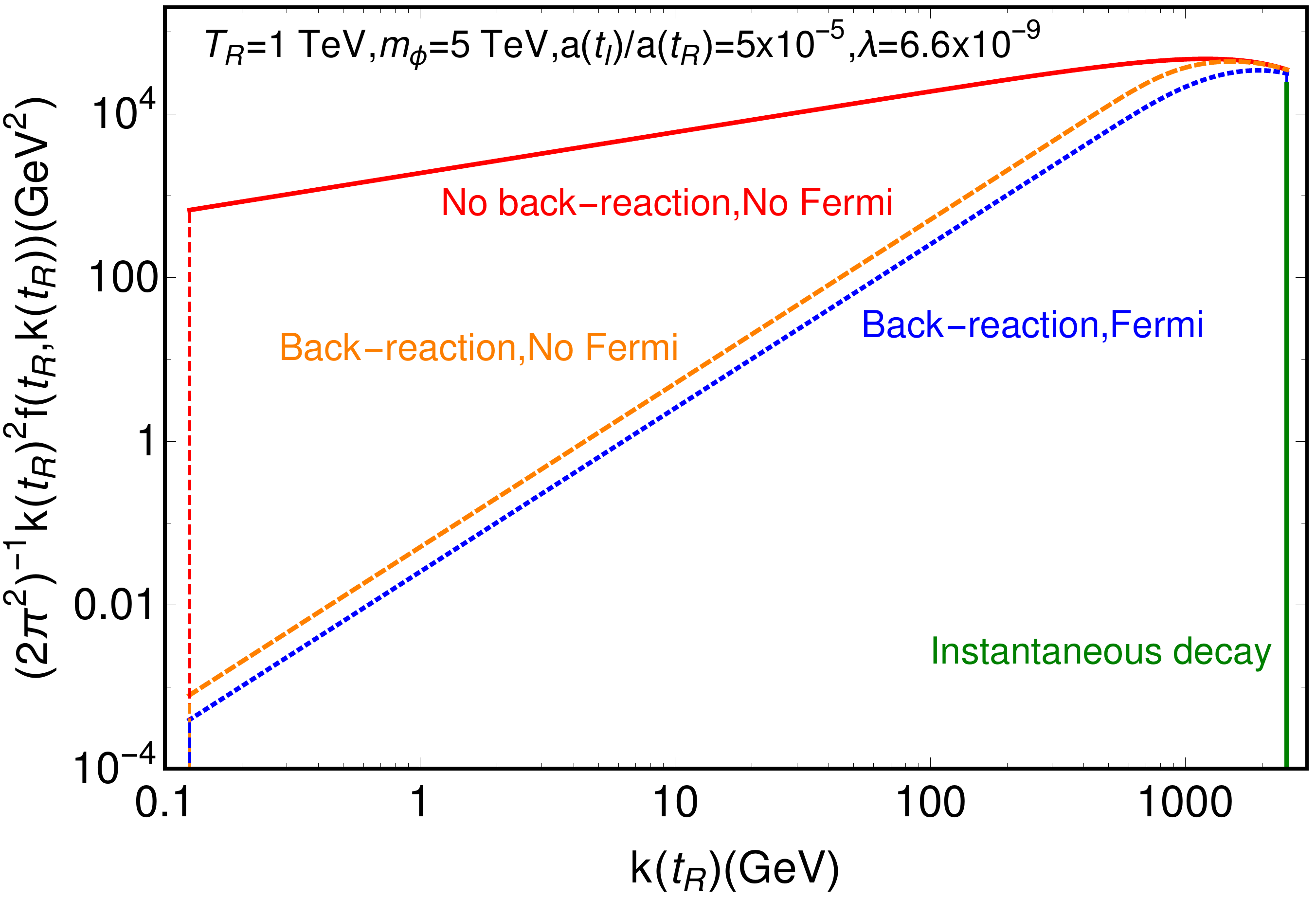} \hspace{0.0cm}
\includegraphics[scale=0.265]{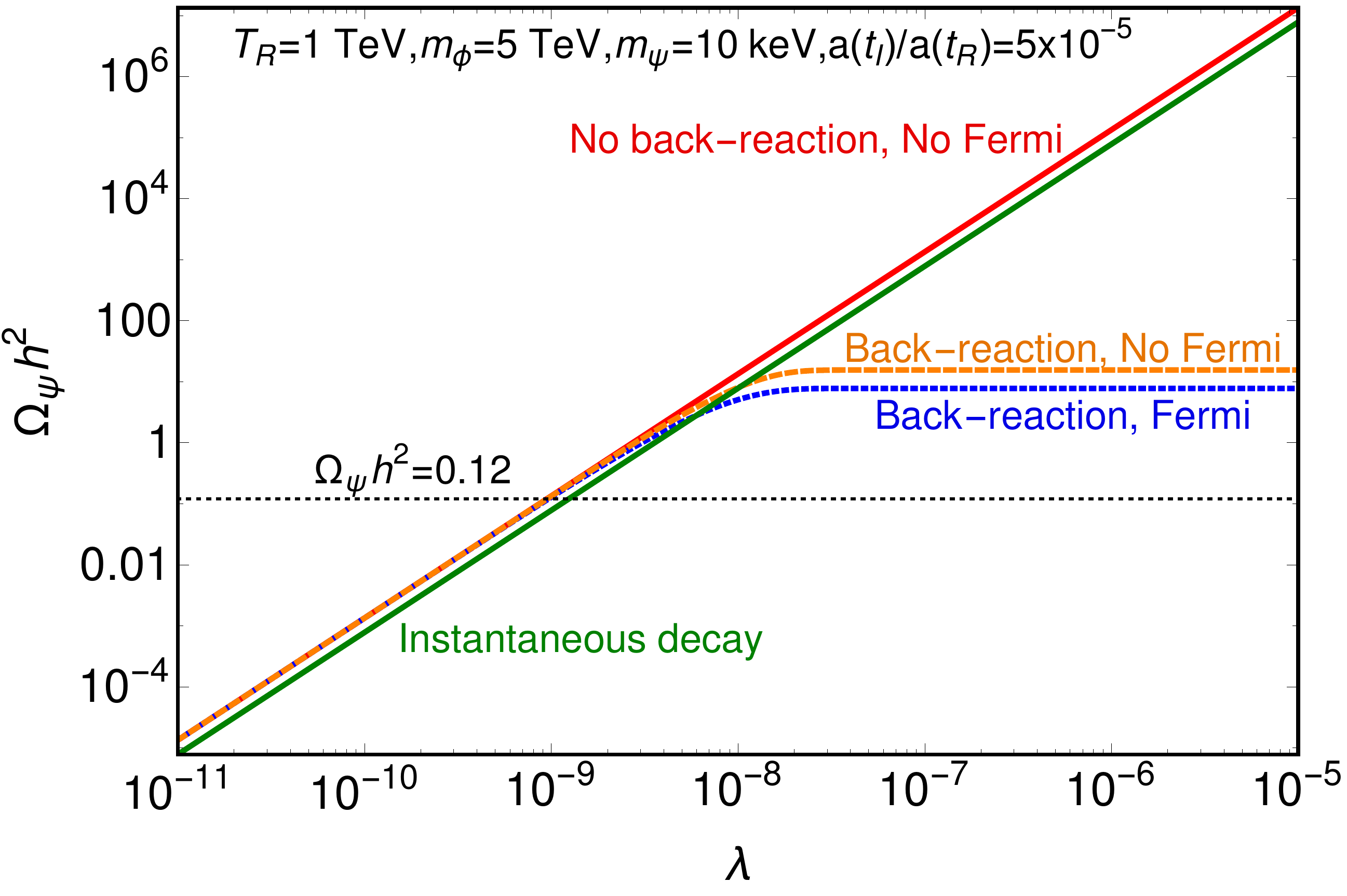}
\end{center}
\caption{\small{\it{{\bf Left panel:} DM momentum distribution at the reheating time $t_R \sim 1/\Gamma_\phi$, (a) without the effects of back-reaction or Fermi statistics (red solid curve), (b) including back-reaction but with classical statistics (orange dashed curve) and (c) including both back-reaction and Fermi statistics (blue dotted curve). For comparison we also show the distribution with the (d) instantaneous decay approximation (green solid line). {\bf Right panel:} The DM relic abundance $\Omega_\psi h^2$ as a function of the DM-inflaton coupling $\lambda$ in all the four cases (a)-(d).} } }
\label{Fig:mom_num}
\end{figure}
We compare these three distribution functions and the resulting DM number densities in Fig.~\ref{Fig:mom_num}, for fixed values of the relevant parameters shown in the figure. From the left panel of Fig.~\ref{Fig:mom_num}, we  see that scenario-(a) (red solid curve), for which a closed form analytic expression for $n_\psi(t_R)$ can be obtained, overpredicts the DM momentum distribution in all but the highest momentum values. In scenario-(b)(orange dashed curve), the back-reaction effect suppresses the distribution throughout $-$ lower the momentum at reheating, higher the suppression. This is because the particles produced at the earliest times have the lowest momentum at reheating due to a large redshift effect. On the otherhand, it is these particles which have the highest probability of recombining to produce an inflaton through back-reactions, due to the longer timespan available. Including the effect of Fermi statistics in scenario-(c) (blue dotted curve) suppresses the distribution further due to Pauli blocking by a smaller ratio. For comparison, we also show the delta-function-like distribution in (d) the instantaneous decay approximation (green solid line). For our subsequent numerical studies, we have included the effects of both back-reaction and Fermi statistics.

In the right panel of Fig.~\ref{Fig:mom_num} we have shown the DM relic abundance $\Omega_\psi h^2$ as a function of the DM-inflaton coupling in all the four cases (a)-(d). Following the distribution functions, it is straightforward to understand the $\Omega_\psi h^2$ curves. It is interesting to observe that $\Omega_\psi h^2$ reaches a constant value above a certain coupling of around $\lambda \simeq 10^{-8}$, when back-reactions and Fermi statistics effects are included. This is essentially because for couplings larger than this value, the initial DM number density builds up quickly, such that the forward and backward reaction rates become the same, thereby freezing the change in the co-moving DM number density $n_\psi a^3$. We also observe that the instantaneous decay approximation differs from the scenario-(1) prediction by a constant factor (of about 1.72) as seen in the comparison between Eq.~\ref{eqn:numberint1} with Eq.~\ref{eqn:DMno1}. We find that for $T_R=1{~\rm TeV}$ and $m_{\phi}=5{~\rm TeV}$, a $10{~\rm keV}$ DM particle achieves the relic density of $\Omega_\psi h^2 = 0.12$ for a DM-inflaton coupling of around $\lambda \simeq 10^{-9}$.

\section{Dark matter momentum distribution: effect of inflaton mediated scatterings}
\label{sec:sec3}
The collision term $f_\psi^{\rm coll}(\vec{k},t)$ in the kinetic equation~\ref{eqn:BEinf1} for $f_\psi(\vec{k},t)$ receives further contributions from inflaton-mediated scattering processes, which are necessarily present in the reheating scenario studied here. In particular, the interaction Lagrangian in Eq.~\ref{eqn:lag} implies the following reactions:
\begin{enumerate}
\item  $s-$channel and $t-$channel inflaton mediated scatterings: $\psi(p_1) \overline{\psi}(p_2) \leftrightarrow \psi(p_3) \overline{\psi}(p_4)$
\item $t-$channel inflaton mediated scattering: $\psi(p_1) h(p_2) \leftrightarrow \psi(p_3) h(p_4)$, and its CP-conjugate process  $\overline{\psi}(p_1) h(p_2) \leftrightarrow \overline{\psi}(p_3) h(p_4)$
\item  $s-$channel inflaton mediated scattering: $\psi(p_1) \overline{\psi}(p_2) \leftrightarrow h(p_3) h(p_4)$.
\end{enumerate}
Here, we have assumed the reheating temperature $T_R>m_h$, so that the Higgs boson $h$ is present in the thermal bath. For $T_R<m_h$, the same interaction Lagrangian Eq.~\ref{eqn:lag} would imply DM scatterings with $W-$bosons, or $b-$quarks or SM leptons, depending upon the relevant temperature. 

In presence of the above $2 \rightarrow 2$ scatterings, the DM momentum distribution evolves according to the following Boltzmann equation
\begin{equation}
\frac{\partial\,f(E_1,t)}{\partial\,t} - H\,|\vec{p}_1|\,\frac{\partial f(E_1,t)}{\partial\,|\vec{p}_1|} = C(E_1,t),
\label{eqn:BEeqscatt}
\end{equation}
where we have written the Liouville operator in the LHS in polar co-ordinates for the momentum $\vec{p}_1$, by imposing the condition of isotropy to set the derivatives with respect to the polar and azimuthal angles to zero. The collision term $C(E_1,t)$ encapsulates the effects of the $2 \rightarrow 2$ processes given above and can be written as follows:
\begin{eqnarray}
C(E_1,t) &=& -\frac{1}{2\,E_1}\int\,\frac{d^3\,p_2}{(2\pi)^3\,2E_2}\int\,\frac{d^3\,p_3}{(2\pi)^3\,2E_3}\,\int\,\frac{d^3\,p_4}{(2\pi)^3\,2E_4}\,\nonumber\\
&&(2\pi)^4\,\delta^4(p_1+p_2-p_3-p_4)|M|^2\,(f(E_1,t)f(E_2,t)-f(E_3,t)f(E_4,t)),
\label{eqn:colltermscatt}
\end{eqnarray}
which can be conveniently broken up into two terms,
\begin{eqnarray}
\label{eqn:colltermscatt1}
C^{\rm BW}(E_1,t) &=& \frac{1}{2\,E_1}\int\,\frac{d^3\,p_2}{(2\pi)^3\,2E_2}\int\,\frac{d^3\,p_3}{(2\pi)^3\,2E_3}\,\int\,\frac{d^3\,p_4}{(2\pi)^3\,2E_4}\nonumber\\
&& (2\pi)^4\,\delta^4(p_1+p_2-p_3-p_4)|M|^2\,f(E_3,t)f(E_4,t),\\
C^{\rm FW}(E_1,t) &=& \frac{1}{2\,E_1}\int\,\frac{d^3\,p_2}{(2\pi)^3\,2E_2}\int\,\frac{d^3\,p_3}{(2\pi)^3\,2E_3}\,\int\,\frac{d^3\,p_4}{(2\pi)^3\,2E_4}\nonumber\\
&&(2\pi)^4\,\delta^4(p_1+p_2-p_3-p_4)|M|^2\,f(E_1,t)f(E_2,t).\nonumber\\
\label{eqn:colltermscatt2}
\end{eqnarray}
In contrast to the collision term for the $\phi \leftrightarrow \psi \overline{\psi}$ process, in which the effect of Pauli blocking was included while determining the DM momentum distribution at reheating in Sec.~\ref{sec:sec2}, for the $2 \rightarrow 2$ processes considered above, we have made the approximation of omitting the effects of quantum statistics. This is a small effect here, and its inclusion makes the numerical evaluation of the collision integral more computationally intensive. 

For the backward term in Eq.~\ref{eqn:colltermscatt1}, we can integrate the momentum $p_2$ using the momentum conserving delta-function and then parametrize rest of the momenta as follows~\cite{Hannestad:1995rs,Kreisch:2019yzn,Ala-Mattinen:2022nuj,Du:2021jcj}:
\begin{equation}
p_1 = (E_1,0,0,p_1),\,p_3 = (E_3,0,p_3\sin\theta,p_3\cos\theta),\,p_4 = (E_4,p_4\sin\alpha\sin\beta,p_4\sin\alpha\cos\beta,p_4\cos\alpha).
\label{eqn:mtmBW}
\end{equation}
With this, we can reduce the backward collision term to the following form
\begin{equation}
C^{\rm BW}(E_1,t) = \frac{1}{128\pi^4\,E_1}\int\frac{p^2_3\,d\,p_3}{\sqrt{p^2_3+m^2_3}}\,\int\frac{p^2_4\,d\,p_4}{\sqrt{p^2_4+m^2_4}}\,f(E_3,t)f(E_4,t)\,F(p_1,p_3,p_4)\Theta(p_3+p_4-p_1),
\label{eqn:colltermscatt1A}
\end{equation}
where, the angular integral is given by
\begin{equation}
F(p_1,p_3,p_4) =  2\int_{-1}^{1} d\cos\theta \int_{-1}^{1} d\cos\alpha \,\left[\frac{|M|^2}{\sqrt{a\,\cos^2\alpha + b\,\cos\alpha + c}} \Theta(a\,\cos^2\alpha + b\,\cos\alpha + c)\right].
\label{eqn:colltermBWF}
\end{equation}
In Eqn.~\ref{eqn:colltermBWF} the parameters $a,b$ and $c$ are defined as follows:
\begin{eqnarray}
a &=& 4\,p^2_4\,(-p^2_1-p^2_3+2\,p_1\,p_3\,\cos\theta),\\
b &=& -4\,p_4\,(p_1-p_3\cos\theta)\,(m^2_1 - m^2_2 + m^2_3 + m^2_4+2(E_3 E_4  - E_1 E_3 - E_1 E_4)+2\,p_1\,p_3\cos\theta)\nonumber\\\\
c &=& 4\,p^2_3\,p^2_4\,(1-\cos^2\theta) - (m^2_1 - m^2_2 + m^2_3 + m^2_4+2(E_3 E_4  - E_1 E_3 - E_1 E_4)+2\,p_1\,p_3\cos\theta)^2,\nonumber\\
\label{eqn:BWparams}
\end{eqnarray}
with $E_i=\sqrt{p^2_i+m^2_i}$. The Heaviside theta 
functions in Eqs.~\ref{eqn:colltermscatt1A} and~\ref{eqn:colltermBWF} put the relevant kinematic  
constraints on the phase-space variables. 

Similarly, for the forward collision term in Eq.~\ref{eqn:colltermscatt2}, we can integrate the momentum $p_4$ and 
parametrize the other three momenta as follows:
\begin{equation}
p_1 = (E_1,0,0,p_1), p_3 = (E_3,0,p_3 \sin\theta, p_3 \cos\theta), p_2 = (E_2, p_2 \sin\alpha\sin\beta,p_2 \sin\alpha\cos\beta, p_2\cos\alpha).
\label{eqn:mtmFW}
\end{equation}
Using these parametrizations, we obtain the following form of the forward collision term
\begin{equation}
C^{\rm FW}(E_1,t) = \frac{1}{128\pi^4\,E_1}\int\,\frac{p^2_2\,dp_2}{\sqrt{p^2_2+m^2_2}}\int\,\frac{p^2_3\,d\,p_3}{\sqrt{p^2_3+m^2_3}}\,f(E_1,t)f(E_2,t) F^\prime(p_1,p_2,p_3)\Theta(p_1+p_2-p_3),
\label{eqn:colltermscatt2A}
\end{equation}
with the angular integral given by
\begin{equation}
F^\prime(p_1,p_2,p_3) = 2 \int_{-1}^{1} d\cos\theta \int_{-1}^{1} d\cos\alpha \left[ \frac{|M|^2}{\sqrt{a^{\prime}\cos^{2}\alpha + b^{\prime}\cos\alpha + c^\prime}} \Theta(a^{\prime}\cos^{2}\alpha + b^{\prime}\cos\alpha + c^\prime) \right].
\label{eqn:colltermFWF}
\end{equation}
Here again the relevant parameters are defined in terms of the momenta as follows
\begin{eqnarray}
a^\prime &=& 4\,p^2_2\,(-p^2_1-p^2_3+2\,p_1\,p_3\,\cos\theta),\\
b^\prime &=& 4\,p_2\,(p_1-p_3\cos\theta)\,(m^2_1 + m^2_2 + m^2_3 - m^2_4+2(E_1 E_2  - E_1 E_3 - E_2 E_3)+2\,p_1\,p_3\cos\theta)\nonumber\\\\
c^\prime &=&  4\,p^2_3\,p^2_2\,(1-\cos^2\theta) - (m^2_1 + m^2_2 + m^2_3 - m^2_4+2(E_1 E_2  - E_1 E_3 - E_2 E_3)+2\,p_1\,p_3\cos\theta)^2,\nonumber\\
\label{eqn:FWparams}
\end{eqnarray}
with $E_i=\sqrt{p^2_i+m^2_i}$.

The matrix elements squared appearing in Eqs.~\ref{eqn:colltermBWF} and \ref{eqn:colltermFWF} are obtained to be the following
\begin{enumerate}
	\item $\psi \overline{\psi}\rightarrow \psi \overline{\psi}$:
	\begin{eqnarray}
	|M|^2 &=&  4\lambda^4  \frac{(s-4m^2_\psi)^2}{(s-m^2_\phi)^2+m^2_\phi\Gamma^2_\phi} + 4\lambda^4 \frac{(4m^2_\psi-t)^2}{(t-m^2_\phi)^2} 
	\end{eqnarray}
	\item $\psi h \rightarrow \psi h$ or $ \overline{\psi} h \rightarrow  \overline{\psi} h$:
	\begin{eqnarray}
	|M|^2 &=& 2\lambda^2 \mu^2_H \frac{4m^2_\psi-t}{(t-m^2_\phi)^2}
	\end{eqnarray}
	\item $\psi  \overline{\psi}\rightarrow h h$: 
	\begin{eqnarray}
	|M|^2 &=& 2\lambda^2 \mu^2_H \frac{s-4m^2_\psi}{(s-m^2_\phi)^2+m^2_\phi\Gamma^2_\phi}
	\end{eqnarray}
\end{enumerate}
where $s=(p_1+p_2)^2=(p_3+p_4)^2$ and $t=(p_3-p_1)^2$ represent the standard Mandelstam variables.

We numerically solve the Boltzmann Eq.~\ref{eqn:BEeqscatt} with the collision term given in Eq.~\ref{eqn:colltermscatt}. The angular integrals in Eqs.~\ref{eqn:colltermBWF} and \ref{eqn:colltermFWF} are evaluated using Monte-Carlo integration implemented in $\texttt{CUBA}$~\cite{Hahn:2004fe} to calculate the functions $F(p_1,p_3,p_4)$ and $F^\prime(p_1,p_2,p_3)$, respectively. These functions are then fed into Eqs.~\ref{eqn:colltermscatt1A} and \ref{eqn:colltermscatt2A} to obtain the full collision integral, which are subsequently used to solve Eq.~\ref{eqn:BEeqscatt}. 

We have followed the evolution of the DM momentum distribution from reheating ($t_R$) to  
matter-radiation equality ($t_{\rm EQ}$). The boundary condition is set by the 
distribution function at $t=t_R$ which is shown in Eq.~\ref{eqn:distFB}. 
For numerical convenience we have traded the time variable in terms of the temperature of the 
SM plasma using the following relation:
\begin{equation}
H = \frac{1}{2t} = \sqrt{\frac{\pi^2\,g_*(T)}{90}}\,\frac{T^2}{M_{\rm Pl}}.
\label{eqn:Hubbleraddom}
\end{equation} 
The Boltzmann equation is then solved using the backward difference formula in $200$ bins spanning the momentum range $10^{-10}$ GeV to $10^{10}$ GeV.
To solve the Boltzmann equation we have used a suitable differential equation solver from the $\texttt{SciPy}$ library of $\texttt{Python}$~\cite{Ala-Mattinen:2022nuj,Du:2021jcj}. We found that for all the cases considered in this work, the shape of the distribution freezes around $T \sim 100\,{\rm GeV}$, and then subsequently redshifts till matter-radiation equality, implying that the scattering processes are no longer active below this bath temperature.

\section{Dark matter average velocity and free-streaming length}
\label{sec:sec4}
Having obtained the momentum distribution of the DM particles, including the effects of inflaton-mediated scattering processes, we can easily estimate the average DM velocity at different epochs, and from there the DM free-streaming length $\lambda_{\rm FSH}$. We recall that until the DM perturbations become Jeans unstable and begin to grow at $t=t_{\rm EQ}$, $t_{\rm EQ}$ being the time of matter-radiation equality, collisionless particles can stream out of overdense regions into underdense regions. This process can smooth out inhomogeneities, and would be constrained by the DM power spectrum deduced from cosmological observations~\cite{Kolb:1990vq}. This effect can also be approximately understood using the DM co-moving free-streaming length.

The co-moving free-streaming length is defined as the co-moving distance travelled by the DM particles between the time of decoupling of scattering reactions $t_{\rm dec}$ and the time of matter-radiation equality $t_{\rm EQ}$~\cite{Kolb:1990vq,Boyarsky:2008xj}:
\begin{eqnarray}
\lambda_{\rm FSH} = \int_{t_{\rm dec}}^{t_{\rm EQ}} \frac{\langle v(t) \rangle}{a(t)}dt,
\label{eqn:FSHlength}
\end{eqnarray}
where, the average velocity at the time $t$, $\langle v(t) \rangle$ is given by
\begin{eqnarray}
\langle v(t) \rangle &=& \left({\int \frac{d^3\vec{p}}{(2\pi)^3} \dfrac{p}{\sqrt{p^2+m^2_\psi}} f(p,t)}\right)\bigg/\left({\int \frac{d^3\vec{p}}{(2\pi)^3} f(p,t)}\right),
\label{eqn:vav1}
\end{eqnarray}
Eq.~\ref{eqn:FSHlength} can be re-written by exchanging time $t$ with the scale factor $a(t)$ as
\begin{eqnarray}
\lambda_{\rm FSH} = \int_{a_{\rm dec}}^{a_{\rm EQ}} \frac{\langle v(a) \rangle}{a^2\,H(a)}da.
\label{eqn:FSHlength2}
\end{eqnarray}

If the DM decoupling takes place in a radiation-dominated Universe, then, we have the following relations between the scale factor and temperature: $a_{\rm dec}= \dfrac{T_0}{T_{\rm dec}}\left(\dfrac{g_{*,s}(T_0)}{g_{*,s}(T)}\right)^{1/3}$ and $a_{\rm EQ}= \dfrac{T_0}{T_{\rm EQ}}$, as the relativistic degrees of freedom $g_{*,s}$ at equality and the present epoch are the same. Here $T_0$ denotes the present temperature, with $a(T_0)=1$. Including the contributions from matter and radiation energy densities, the Hubble parameter may be expressed as
\begin{eqnarray}
H(a) &=& H_0\sqrt{\Omega_R}\frac{\sqrt{ 1 + a/a_{\rm EQ}}}{a^2},
\label{eqn:Hubscale}	
\end{eqnarray}
where, $H_0$ is the Hubble expansion parameter at the present epoch, and $\Omega_R = \rho_R(T_0)/\rho_c$, $\rho_c$ being the critical density. Using Eq.~\ref{eqn:Hubscale}, the free-streaming length can be expressed as
\begin{eqnarray}
\lambda_{\rm FSH} &=& \frac{a^2_{\rm EQ}}{H_0\sqrt{\Omega_R}}\int_{a_{\rm dec}}^{a_{\rm EQ}} \frac{\langle v(a) \rangle}{a^{3/2}_{\rm EQ}\sqrt{a+a_{\rm EQ}}}da.
\label{eqn:FSHlength3}
\end{eqnarray}    

On using the central numerical values from the Planck data~\cite{Planck:2018vyg}, with the parameters $H_0 = 67.66\,{\rm km}\,{\rm s}^{-1}\,{\rm Mpc}^{-1}$, $\Omega_M  = 0.1424\,h^{-2}$~\cite{Planck:2018vyg} and $\Omega_R  = 4.15\times 10^{-5}\,h^{-2}$~\cite{Weinberg:2008zzc}, we obtain,
\begin{eqnarray}
\lambda_{\rm FSH} &\simeq& 0.0394\,{\rm Mpc}\,\int_{a_{\rm dec}}^{a_{\rm EQ}} \frac{\langle v(a) \rangle}{a^{3/2}_{\rm EQ}\sqrt{a+a_{\rm EQ}}}da,
\label{eqn:FSHlength4}
\end{eqnarray} 
where $a_{\rm EQ}=\Omega_R/\Omega_M = 2.91\times 10^{-4}$. 

It is convenient to re-write the momentum integral for $\langle v(a) \rangle$ in terms of the scaled variable $q = \dfrac{p_{\rm dec}}{T_{\rm dec}} =  \dfrac{p}{T}\left(\dfrac{g_{*,s}(T_{\rm dec})}{g_{*,s}(T)}\right)^{1/3} = a \dfrac{p}{T_0}\left(\dfrac{g_{*,s}(T_{\rm dec})}{g_{*,s}(T_0)}\right)^{1/3} $, leading to the expression
\begin{eqnarray}
\langle v(a) \rangle &=& \left({\int dq \frac{q^3}{\sqrt{q^2+\frac{m^2_\psi\,a^2}{T_0^2}\left(\frac{g_{*}(T_{\rm dec})}{g_{*}(T_0)}\right)^{2/3}}} f(q,T)}\right)\bigg / \left({\int dq q^2\,f(q,T)}\right).
\label{eqn:vav2}
\end{eqnarray}

\section{Results}
\label{sec:sec5}
The $2 \rightarrow 2$ scattering cross-sections discussed in the previous section are functions of the DM-inflaton coupling $\lambda$, the SM-inflaton coupling $\mu_H$, the inflaton mass $m_\phi$ and the DM mass $m_\psi$. Since the inflaton decay to the SM sector dominates its total decay width, we can trade the coupling $\mu_H$ with the reheating temperature $T_R$. Thus, for a fixed $m_\psi$ and $\lambda$, the scattering rates are controlled by $T_R$ and $m_\phi$. $T_R$ also controls the branching fraction of inflaton decay to DM, and hence its number density at reheating. In addition to these effects, $T_R$ and $m_\phi$ affect the impact of scatterings on the DM momentum distribution in a number of ways, as discussed subsequently. Since the collision integrals and the solution of the Boltzmann equations for the DM momentum distribution are numerically highly demanding even for a single set of parameter values, a full scan of the $\{T_R, m_\phi, m_\psi, \lambda \}$ parameter space is beyond the scope of this work. We shall therefore choose a particular set of these parameters to illustrate the effect of collisions.

The effect of the scatterings is also controlled by how broad the DM momentum distribution is at reheating. This in turn is dependent on the duration of the reheating process. We quantify the duration of the reheating process by the scale factor ratio $a(t_I)/a(t_R)$, where $t_I$ denotes the beginning of the reheating phase, and $t_R$ denotes the beginning of radiation domination with $\Gamma_\phi t_R \sim 1$. Smaller values of $a(t_I)/a(t_R)$ would correspond to a longer duration of the reheating process.

\begin{figure} [htb!]
\begin{center} 
\includegraphics[scale=0.4]{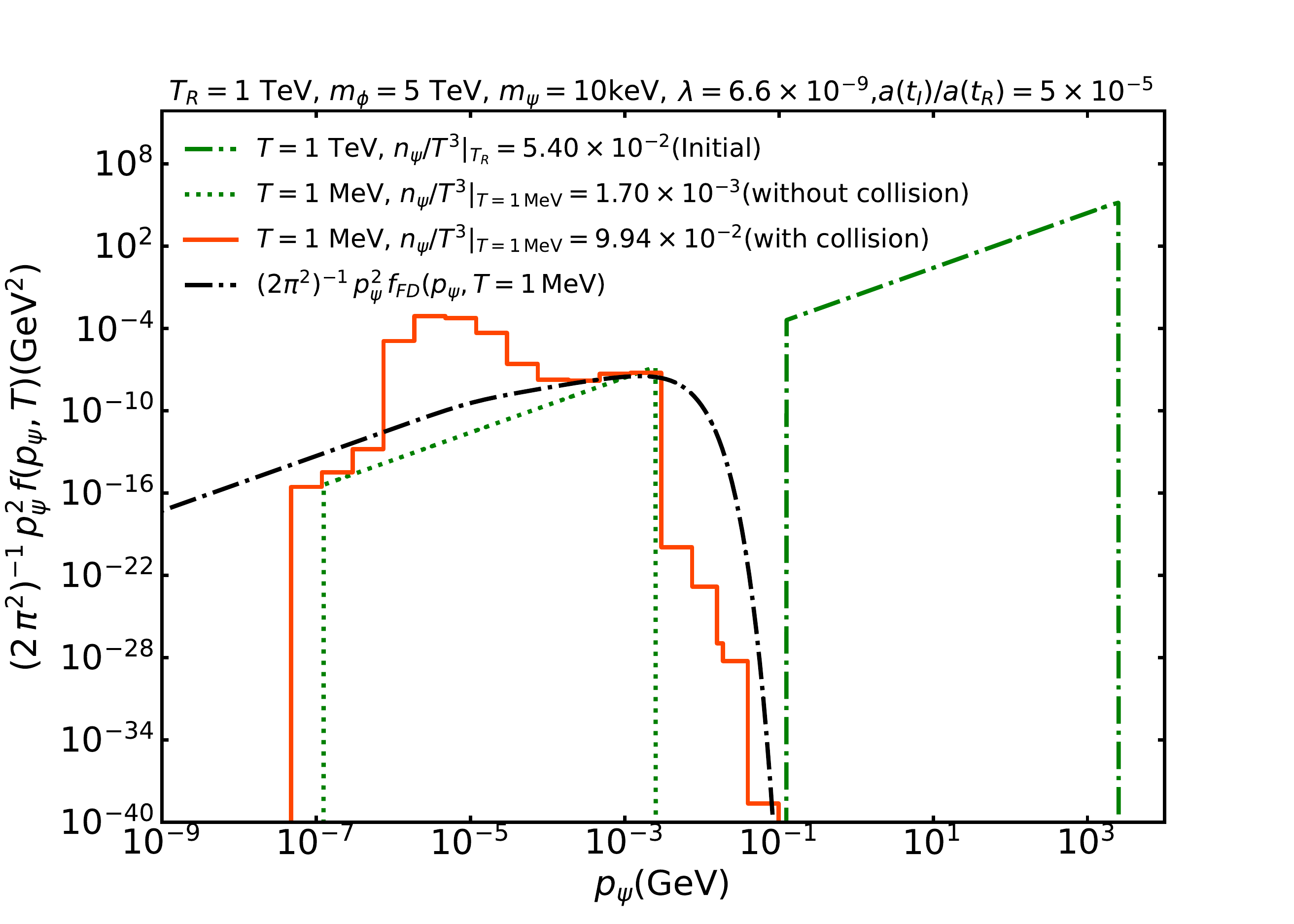} \hspace{0.0cm}
\end{center}
\caption{\small{\it{Momentum distribution of DM including the effect of $2 \rightarrow 2$ inflaton-mediated scattering processes (red solid histogram), at the epoch with SM bath temperature $T=1{~\rm MeV}$. For comparison, we also show the distribution at reheating with $T_R = 1 {~\rm TeV}$ (green dot-dashed curve), and the corresponding red-shifted distribution without the effect of collisions at $T=1{~\rm MeV}$ (green dotted curve). Also shown is the thermal Fermi-Dirac distribution $f_{\rm FD}$ at $T=1{~\rm MeV}$ for a particle of mass $10{~\rm keV}$ with zero chemical potential (black dot-dashed curve).} } }
\label{Fig:dist1}
\end{figure}
In Fig.~\ref{Fig:dist1}, we show the momentum distribution of DM including the effect of the three $2 \rightarrow 2$ inflaton-mediated scattering processes discussed in Sec.~\ref{sec:sec3} (red solid histogram). We have shown the distribution at the epoch with SM bath temperature $T=1{~\rm MeV}$. The corresponding distribution at the matter-radiation equality, $T_{\rm EQ} \sim 1{~\rm eV}$ remains the same in shape, with each momentum getting red-shifted by the same factor.  For comparison, we also show the distribution at reheating with $T_R = 1 {~\rm TeV}$ (green dot-dashed curve) obtained earlier, and the corresponding red-shifted distribution without the effect of collisions at $T=1{~\rm MeV}$ (green dotted curve). As we can see by comparing the distributions at $T=1{~\rm MeV}$ with and without collisions, the two-body scatterings lead to a significant modification over the whole momentum range. However, the effective change is more pronounced in the lower and medium momentum range. The modifications are driven by two factors $-$ one being the energy exchange through the DM self-scatterings and the $t-$channel scattering of DM particles with the SM bath, and the other being the non-thermal freeze-in production of DM through the inflaton-mediated s-channel $h(p_3) h(p_4) \rightarrow  \psi(p_1) \overline{\psi}(p_2)$ process. For the DM coupling and inflaton mass chosen, the collisions do not thermalize the DM with the SM sector. In order to demonstrate this, we also show in Fig.~\ref{Fig:dist1} the thermal Fermi-Dirac distribution at $T=1{~\rm MeV}$ for a particle of mass $10{~\rm keV}$ with zero chemical potential (black dot-dashed curve). Clearly, in addition to the small difference in the normalization, the shape of the distributions, especially at lower momenta, differ significantly. All the results are shown with the inflaton mass $m_\phi = 5\,$TeV, reheating temperature $T_{R} = 1\,$TeV, DM mass $m_\psi=10{~\rm keV}$, DM-inflaton coupling $\lambda=6.6 \times 10^{-9}$ and the scale factor ratio chosen as $a(t_I)/a(t_R) = 5 \times 10^{-5}$.

\begin{figure} [htb!]
\begin{center} 
\includegraphics[scale=0.35]{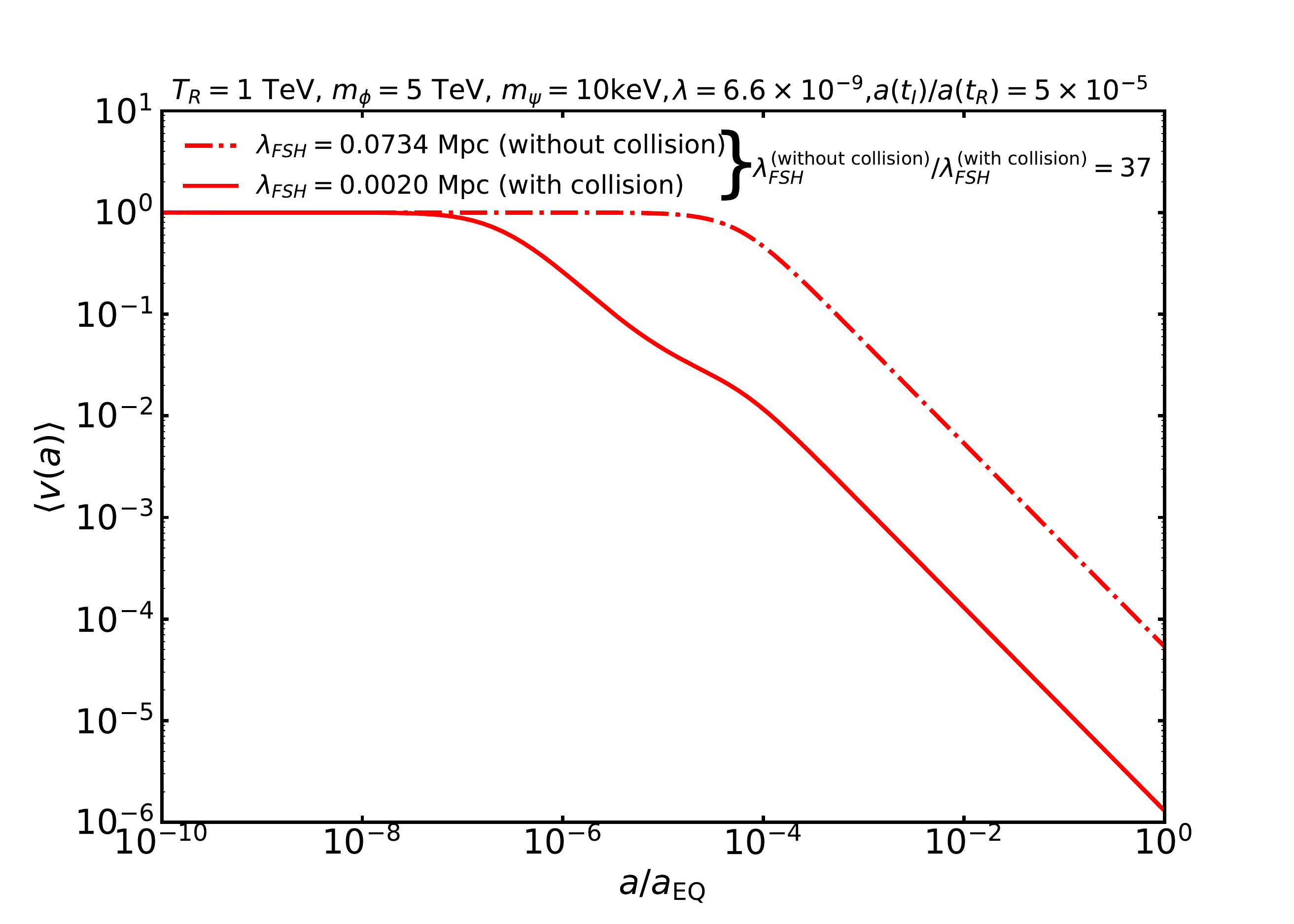} \hspace{0.0cm}
\end{center}
\caption{\small{\it{The average DM velocity $\langle v \rangle$ as a function of the scale factor ratio $a/a_{\rm EQ}$, with the effect of the inflaton-mediated $2 \rightarrow 2$ scatterings included (solid curve) and without them (dot-dashed curve). All the relevant parameters are kept fixed at the values shown in the figure, which are the same as in Fig.~\ref{Fig:dist1}. The value of the DM free-streaming length $\lambda_{\rm FSH}$ in both cases are also shown, which, along with $\langle v \rangle$ is significantly modified by the effect of collisions.} } }
\label{Fig:velocity}
\end{figure}
The impact of the scattering processes can be quantified using the observationally relevant quantities such as the average DM velocity discussed in Sec.~\ref{sec:sec4}. To understand this effect, we show in Fig.~\ref{Fig:velocity}, $\langle v \rangle$ as a function of the scale factor ratio $a/a_{\rm EQ}$, with the effect of the inflaton-mediated $2 \rightarrow 2$ scatterings included (solid curve) and without them (dot-dashed curve). All the relevant parameters are kept fixed at the values shown in Fig.~\ref{Fig:velocity}, which are the same as in Fig.~\ref{Fig:dist1}. There are two different regimes in both the curves. The first is the relativistic regime, in which $\langle v \rangle \simeq 1$. For higher values of the scale factor we have the non-relativistic regime in which we observe a constant slope for the curve without collisions, as in this case, $\langle v \rangle \propto \langle |\vec{p}| \rangle \propto 1/a$. The onset of the non-relativistic regime is found to be much earlier for the case with scatterings included. This is because, the scattering processes lead to the enhanced occupation of lower momentum modes, which in turn reduces the average velocity. As we can see from this figure, the difference between the two cases is highly significant, with the $\langle v \rangle$ at matter-radiation equality, where $a/a_{\rm EQ}=1$, differing by a factor of $40$. This effect can also be quantified by the free-streaming length of DM defined in Sec.~\ref{sec:sec4}, which differs in the two cases by a factor of around $37$. Thus for the DM mass shown, $m_\psi = 10~{\rm keV}$, while the value of $\lambda_{\rm FSH}$ without including the collisions is found to be $0.0734~{\rm Mpc}$, inclusion of the collision effects reduces it to $0.0020~{\rm Mpc}$. Therefore within this estimate, while the number without the effect of collision would imply strong constraints on such a DM from the Lyman$-\alpha$ forest data~\cite{Irsic:2017ixq}, including the collision effects shows that for fixed values of the other parameters, this DM mass is highly consistent with the structure formation requirements. This crucial effect of the scattering processes is the primary result of this paper.

\begin{figure} [htb!]
\begin{center} 
\includegraphics[scale=0.4]{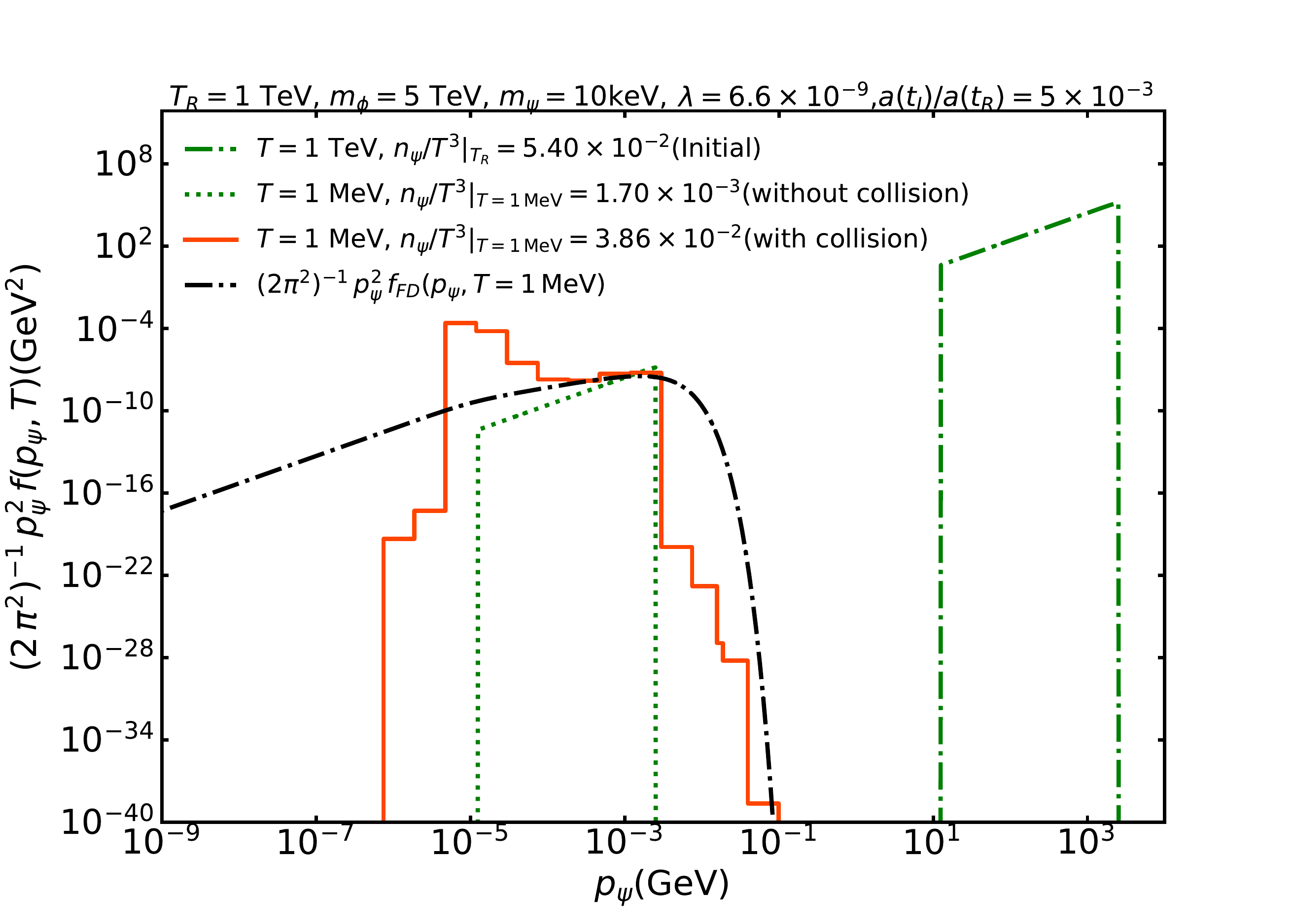} \hspace{0.0cm}
\end{center}
\caption{\small{\it{Same as Fig.~\ref{Fig:dist1}, with a modified $a(t_I)/a(t_R) = 5 \times 10^{-3}$, where the scale factor ratio signifies the duration of the reheating process.} } }
\label{Fig:dist2}
\end{figure}
We observe that for fixed values of the couplings, the scattering effects are sensitive to the duration of the reheating process. This is because of two reasons. Firstly, as seen earlier, for example in Eq.~\ref{eqn:numberint1}, the duration of the reheating process affects the number density of DM produced at reheating, which in turn enters the reaction rates. Secondly, longer the duration, the momentum distribution at reheating is broader and more stretched towards lower momentum values. A larger range of DM momenta increases the probability of the scatterings further. Even if the number density remains very similar, the broadening of the distribution itself can make a large difference in the effect of the collisions.  In order to demonstrate this effect, we show in Fig.~\ref{Fig:dist2} the DM momentum distributions with $a(t_I)/a(t_R) = 5 \times 10^{-3}$. For this figure, all other parameters have been kept fixed at the same values as in Fig.~\ref{Fig:dist1}, for which $a(t_I)/a(t_R)$ was taken to be  $5 \times 10^{-5}$.  For the two $a(t_I)/a(t_R)$ values shown here, the DM number density at $T_R$ are nearly the same. However, by comparing Fig.~\ref{Fig:dist2} with Fig.~\ref{Fig:dist1}, we see that larger the duration of reheating, the initial distribution at $T_R$ is broader, which is easily understood by taking into account the larger redshift of the DM momenta the earlier it is produced. By further comparison of Fig.~\ref{Fig:dist2} with Fig.~\ref{Fig:dist1}, we find that the effective impact of the collisions on the distribution is larger for a smaller $a(t_I)/a(t_R)$.

\begin{table}[htb!]
\footnotesize
\begin{center}
\begin{tabular}{|c|ccc|ccc|}
\hline
&  \multicolumn{3}{c}{Without Collision} \vline & \multicolumn{3}{c}{With Collision}\vline\\
$a(t_I)/a(t_R)$  & $\langle v(a_{\rm EQ}) \rangle$ & $\lambda_{\rm FSH}$ (Mpc) & $\frac{n_\psi(T_0)}{T^3_0}$  & $\langle v(a_{\rm EQ})\rangle$ & $\lambda_{\rm FSH}$ (Mpc) & $\frac{n_\psi(T_0)}{T^3_0}$ \\
\hline 	
\hline
$5\times 10^{-1}$& $5.56\times 10^{-5}$ & 0.0763 & $1.26 \times 10^{-3}$ & $5.51 \times 10^{-5}$ & 0.0756 & $1.38 \times 10^{-3}$ \\
\hline
$5\times 10^{-3}$& $5.33\times 10^{-5}$ & 0.0733 & $1.70 \times 10^{-3}$ & $3.13 \times 10^{-6}$ & 0.0046 & $3.86 \times 10^{-2}$ \\
\hline
$5\times 10^{-5}$ & $5.33\times 10^{-5}$ & 0.0733 & $1.70 \times 10^{-3}$ & $1.29 \times 10^{-6}$ & 0.0020 & $9.94 \times 10^{-2}$ \\
\hline
\end{tabular}
\end{center}
\caption{\small{\it The average DM velocity at matter-radiation equality ($\langle v(a_{\rm EQ}) \rangle$), the DM free-streaming length ($\lambda_{\rm FSH}$) and the scaled DM number density at the present epoch $n_\psi (T_0)/T^3_0$, without including collision effects  (left column) and with the effect of the inflaton-mediated $2 \rightarrow 2$ scatterings included (right column). All the quantities are shown for three different values of $a(t_I)/a(t_R)$, which signifies the duration of the reheating process. The values of the other relevant parameters are fixed as the inflaton mass $m_\phi = 5\,$TeV, reheating temperature $T_{R} = 1\,$TeV and DM-inflaton coupling $\lambda=6.6 \times 10^{-9}$ and $m_\psi=10{~\rm keV}$, which are the same as in Figs~\ref{Fig:dist1},~\ref{Fig:velocity} and ~\ref{Fig:dist2} }.}
\label{tab:BPchar}	
\end{table}
For a more detailed numerical comparison, we show the average velocity at matter radiation equality ($\langle v(a_{\rm EQ}) \rangle$), the co-moving DM free-streaming length ($\lambda_{\rm FSH}$) and the scaled DM number density at the present epoch $n_\psi (T_0)/T^3_0$, for different values of $a(t_I)/a(t_R) $ in Table~\ref{tab:BPchar}. As we see from this table, these quantities without including the collision effects already converge for $a(t_I)/a(t_R)=5\times 10^{-3}$. This implies that the average properties of the distributions without collisions converge by this value of $a(t_I)/a(t_R)$. However, since the broadness of the distribution affects the collision probabilities significantly, the values of $\langle v(a_{\rm EQ}) \rangle$, $\lambda_{\rm FSH}$ and $n_\psi (T_0)/T^3_0$ are modified by large factors once collision effects are included. $\langle v(a_{\rm EQ}) \rangle$ and $\lambda_{\rm FSH}$ reduce more with longer duration of the reheating phase, since the momenta are distributed more towards smaller values, due to enhanced collision probabilities. On the otherhand, for this same reason, the freeze-in contribution increases, thus enhancing $n_\psi (T_0)/T^3_0$. For our choice of model parameters, with the reheating temperature $T_{\rm R} = 1$ TeV and the inflaton mass $m_{\phi}=5$ TeV, the scaled DM number density increases by upto a factor of around $60$, depending upon the value of the scale factor ratio $a(t_I)/a(t_R)$. For $2T_{\rm R} < m_{\phi}$ as chosen by us, no resonant enhancement of the $h h \rightarrow \bar{\psi} \psi$ process is possible. For the opposite hierarchy of $2T_{\rm R} > m_{\phi}$, this effect can enhance the post-reheating production of DM pair from the primeval plasma substantially, and may thus be disfavoured by relic density considerations.

Given its important role, how small can $r=a(t_I)/a(t_R)$ be? From Eq.~\ref{eqn:infdense}, we find that
\begin{equation}
r^{3} e^{r^{3/2}} \simeq \frac{g^*(T_R) \pi^2 e T_R^4}{\rho_\phi (t_I)},
\end{equation}
where we have used the conditions that the reheating time $t_R \sim 1/\Gamma_\phi$ and $\rho_\phi(T_R) = \rho_R(T_R)$. We have further assumed that between $t_I \leq t \leq t_R$, the Universe is approximately dominated by a stable matter field, and therefore, the scale factor $a(t) \propto t^{2/3}$. Since the constraints from the cosmic microwave background data on inflationary scenarios restrict the energy density in the inflaton field at the end of the slow-roll phase to be around $\rho_\phi (t_I) \lesssim \left(10^{16} {~\rm GeV}\right)^4$, for $T_R=1 {~\rm TeV}$, we have $r \gtrsim 10^{-17}$. If we increase the reheating temperature, the constraint on $r$ becomes stronger. Even though we have chosen $r$ values much larger than the lower bound allowed, as we observe in the above discussion, the physical quantities of interest in the DM sector, such as the average velocity at reheating, its free-streaming length and its density all reach their asymptotic values {\it before collisions} by $r \sim 5 \times 10^{-3}$ for the scenarios studied here, although they differ considerably on inclusion of the collisions.

As we can see from Fig.~\ref{Fig:mom_num}, the DM-inflaton coupling chosen for all the above results, $\lambda=6.6 \times 10^{-9}$, overpredicts the DM relic abundance $\Omega_\psi h^2$ by a factor of $50$. Thus, to be consistent with the measurements of the DM density, such scenarios would require a $50$ times lighter DM particle of mass $m_\psi=0.2{~\rm keV}$. While such a light DM would be ruled out by Lyman-$\alpha$ constraints on its free-streaming length if we did not include the effect of collisions, on including the scattering effects it may still be marginally allowed, with an $\lambda_{\rm FSH} \simeq 0.1{\rm Mpc}$. The same $10{~\rm keV}$ mass DM may also be consistent with the DM density if a small amount of entropy is produced after the decoupling of the collision processes but before the onset of the BBN epoch. We have checked that for the choices of $T_R$ and $m_\phi$ made, if the DM-inflaton coupling is reduced by a factor of around $7$ to match the DM abundance with $m_\psi = 10{~\rm keV}$, the effect of scatterings become less significant, due to the combined effects of the decrease in both the DM initial number density and the scattering cross-sections. This prompts us to vary $T_R$ and $m_\phi$ over a broad range to determine in which region the scatterings become crucial $-$ a numerically very costly task we intend to perform in a future study. Nevertheless, given the analysis presented here, it is clear that the collision effects will be important in a considerable range of the four unknown parameters, being consistent with the DM density requirements.

Can we understand qualitatively the scale of $T_R$ and $m_\phi$ that led to a large effect from scatterings? Since we have considered the effects of on-shell Higgs boson scatterings with the DM particles, the reheating temperature $T_R$ should be larger than the Higgs mass $m_h$. For the given interaction Lagrangian, a lower $T_R$ would also lead to scatterings with $W-$ bosons, $b-$quarks or SM leptons, but with a further suppressed rate. Since the scattering reactions must take place over a length of time for redistributing the momenta considerably, we conclude that $T_R$ should be somewhat higher than $m_h$, thus requiring $T_R$ to be around the TeV scale or higher. 
In most common inflationary scenarios with a reheating phase, we have $T_R<m_\phi$. However, with the momentum scale of the SM particles being of the order of $T_R$, $m_\phi$ cannot be too high compared to $T_R$ for the inflaton-mediated scattering cross-sections to be relevant. Thus with $T_R$ being around a TeV, we can at most have $m_\phi$ to be of the order of a few TeV.

\section{Summary}
\label{sec:sec6}
One of the most well-studied non-thermal mechanism invoked to produce DM in the early Universe is through the decay of the inflaton in the post-inflationary reheating phase. In such scenarios, the resulting DM momentum distribution is taken to be the one produced at reheating, which is then red-shifted at later epochs due to the expansion of the Universe. However, in such scenarios since both the SM and the DM fields couple to the inflaton for successful reheating into these sectors, inflaton mediated $2 \rightarrow 2$ scatterings involving the DM and SM particles are necessarily present. These reactions include the self-scattering of a pair of DM particles, and the $s-$ and $t-$channel inflaton mediated DM-SM scatterings. We address the question of the impact of these scattering processes on the DM momentum distribution in this study. The same parameters that enter the reheating dynamics, and hence the resulting reheating temperature of the SM bath and the density of DM produced, also completely determine the scattering rates. Since the DM momentum distribution directly enters the considerations of structure formation of the Universe, studying the effects of the scatterings is highly relevant observationally as well. This is especially true for light DM, for which constraints coming from, for example, the Lyman-$\alpha$ forest data on its average velocity, and hence the free-streaming properties are stringent. 

We find that for a range of the inflaton mass $m_\phi$ and the reheating temperature $T_R$, including the scattering effects are crucial for determining an accurate DM momentum distribution. For the scenario considered here, in which the inflaton dominantly couples to the SM Higgs boson through renormalizable interactions, this range of $T_R$ and $m_\phi$ is largely controlled by the scale of the Higgs mass $m_h$. In particular, we argue, and find that a reheating temperature $T_R$ at the TeV scale and an inflaton mass scale of a few TeV lead to a large effect from the inflaton mediated $2 \rightarrow 2$ scatterings involving the DM and SM particles. We also show, by studying a number of example cases, that for fixed DM mass, DM-inflaton coupling, $T_R$ and $m_\phi$, the impact of the scattering processes also depend significantly upon the duration of the reheating process. 

We show that the inclusion of the effect of scatterings can modify the average DM velocity at matter-radiation equality and the DM free-streaming length by upto a factor of $40$ for the scenarios studied. This makes including the scattering effects crucial while considering constraints from the Lyman-$\alpha$ forest data. Since the scattering processes tend to redistribute the momenta in such a way that more lower momentum modes are filled, the DM average velocity goes down, and hence the free-streaming length. Therefore, inclusion of the scattering effects makes a much larger range of DM mass allowed from structure formation considerations.

A lot remains to be explored in this direction. First of all, within the scenario studied, a full exploration of the parameter space, being a numerically very challenging task, is left for a future work. As we emphasized above, the scale $T_R$ and $m_\phi$ found to lead to the maximum impact of scatterings is essentially set by the scale of Higgs mass $m_h$. However, this is specific to the interaction Lagrangian considered, which is motivated by the fact that the inflaton has a renormalizable coupling with the Higgs field in the SM. However, in different scenarios for inflation and reheating, the inflaton might have a dominant coupling to other SM fields, thus changing the relevant DM-SM scattering processes. In such a scenario, a different scale of  $T_R$ and $m_\phi$ will lead to the maximum impact of the scattering processes, which should be studied in detail. As shown by the results obtained in this study, precision determination of the momentum distribution of DM produced in inflaton decay including the effect of inflaton mediated scatterings is highly relevant in determining constraints from data on cosmological large-scale structure.
\section*{Acknowledgment}
We thank Deep Ghosh, Sougata Ganguly and Dhiraj Kumar Hazra for helpful discussions.

\end{document}